# SOLEX OBSERVATIONS FOR THE BASS2000 DATABASE, A COLLABORATION PRO-AM

J.-M. Malherbe ; Email : Jean-Marie.Malherbe@obspm.fr ; ORCID id : https://orcid.org/0000-0002-4180-3729

F. Cornu ; Email : Florence.Cornu@obspm.fr

I. Bualé ;   Email: Isabelle.Buale@obspm.fr

Observatoire de Paris, PSL Research University, CNRS, LESIA, Meudon, France

25 May 2023

## ABSTRACT

Systematic observations of the chromosphere and the photosphere started in Meudon Observatory 115 years ago with Deslandres spectroheliograph. An exceptional collection of more than 100 000 monochromatic images in CaII K and Hα spanning more than 10 solar cycles is proposed to the international community by the BASS2000 solar database. We started in 2023 a "PRO-AM" collaboration between professional and amateur astronomers with the Solar Explorer (SOLEX), a compact and high quality spectroheliograph designed by Christian Buil, in order to record images every day, and several times per day, owing to tens of observing stations in various places. This paper summarizes the scientific objectives and provides practical and technical information to amateurs willing to join the observing network.

## INTRODUCTION

The solar atmosphere is composed of three observable layers: the photosphere is the visible one (temperature decreasing from 6000 K to 4500 K in 300 km) with dark magnetic sunspots and bright faculae around. Other layers above, such as the chromosphere (temperature increasing from 4500 K to 8000 K in 2000 km) with dark filaments and bright plages, and the hot corona (2 million K), require spectroscopic means to reveal their structures, respectively via absorption lines of the visible spectrum or emission lines of the ultraviolet spectrum. The solar atmosphere follows a 11-year activity cycle and a 22-year magnetic cycle. Sunspots are regions of intense magnetic fields (2000 Gauss). Bright faculae form, together with sunspots, active regions, and exhibit smaller magnetic fields. Filaments, also called prominences when seen at the limb, are thin, long and high structures of dense material suspended in the corona by weak magnetic fields. Flares and mass ejections occur in active regions a few years around the solar maximum; reconnections convert magnetic energy into kinetic energy (accelerated particles), radiation (X-rays) and heat (brightenings). the maximum of the current cycle (number 25) is forecasted for 2025.

Jules Janssen (1824-1907) founded Meudon observatory in 1875 and introduced physical astronomy in France. He started photographic observations of the solar photosphere and developed high spatial resolution photography revealing thin details such as granules and small sunspots (called pores). Images were recorded on 30 x 30 cm² glass plates. The emulsions were blue-sensitive, in the vicinity of the Fraunhofer G band (4300 Å wavelength). 6000 plates were exposed, but 99 % were unfortunately destroyed or lost. The best observations, showing fantastic details of the granulation, were published in the solar photographic atlas.

Meanwhile, Henri Deslandres (1853-1948) was in charge of organizing a spectroscopic laboratory at Paris Observatory, in the context of the development of physical astronomy initiated by Janssen at Meudon. Deslandres first built a classical spectrograph and studied the line profiles of the ionized Calcium at 3934 Å wavelength (the K line); he succeeded to resolve the fine structure of the core in 1892. This was the starting point of the great adventure of spectroheliographs in France **[ref 1]**. The principle was outlined much earlier by Janssen: with a classical spectrograph, monochromatic images can be delivered by an output slit located in the spectrum (selecting the light of a spectral line) when the input slit scans the solar surface. This is possible by moving either the solar image upon the first slit, or the whole spectrograph. The photographic spectroheliograph was invented on this basis by George Hale in Kenwood (1892) and by Henri Deslandres in Paris (1893), simultaneously but independently. In that respect, Janssen claimed in 1906: *"the method, that I proposed in 1869, consists in the use*

*of a second slit to select in the spectrum, formed by the first slit, a spectral line of interest. Hale and Deslandres have skilfully applied this principle"*. Deslandres moved to meudon in 1898 and developed, with Lucien d'Azambuja (1884-1970) the large quadruple spectroheliograph, operating since 1908. D'Azambuja organized the service of systematic observations of the solar atmosphere, which started at this date in CaII K (K3, the line centre for the chromosphere, and K1v, the violet wing for the photosphere). Hα followed soon in 1909. Observations were interrupted by WW1 but continue today with a modified instrument [ref 2]. The photographic technology was abandoned in 2002 for CCD or CMOS rectangular arrays, implying the suppression of the second slit in the spectrum and the motorized plate translator. Today, Meudon spectroheliograph records full line profiles in 3D FITS files (x, y, λ); the classical images are slices of the data-cube for particular wavelengths. Three lines are observed: Hα (interference order 3) and CaII H and K (which are recorded simultaneously in interference order 5). A standard observation consists of two cubes with short exposure time and two cubes with long exposure time for prominences and an attenuator of neutral density 1 upon the disk. Three series are made daily, weather permitting (250 days/year at maximum).

Long series of continuous observations are of great interest. For instance, series of the CaII K line were collected by the spectroheliographs of Kodaikanal in India, Mount Wilson in the USA, Mitaka in Japan, Sacramento Peak in the USA, Coimbra in Portugal, Meudon in France or Arcetri in Italy. They produced extended archives, some of them covering up to 10 solar cycles, which are convenient to investigate long-term solar activity and study rare events (such as energetic flares, huge sunspot groups or giant filaments). The CaII K intensity is an excellent proxy to reconstruct past irradiance and magnetism, because the line core is greatly enhanced in bright regions (the plages) where the magnetic field and the chromospheric heating are stronger. Series of the Hα line are also very useful for statistical studies concerning filaments and prominences. They are at the base of the synoptic maps which were initiated by d'Azambuja one century ago and produced at Meudon until 2003, for each synodic rotation of the Sun. These synthetic maps provided a good summary, with associated tables, of solar activity. Unfortunately they were stopped due to the lack of manpower.

Spectroheliographs historically produced monochromatic images along solar cycles at the epoch of photographic plates. Most of them observed the CaII K and Hα Fraunhofer lines. They were progressively abandoned in the second half of the twentieth century for telescopes using narrow bandpass filters (such as Fabry-Pérot or Lyot filters), which are more compact and able to observe, at higher cadence, fast evolving events. However, with modern electronic detectors, there is a regain of interest (such as the SOLEX) for imaging spectroscopy, because digital spectroheliographs can now deliver full line profiles for each pixel of the Sun.

The SOLEX instrument, developed recently (2021) by Christian Buil [ref 3] for amateur astronomers, is a compact spectroheliograph with professional capabilities. It offers an outstanding opportunity to develop the survey of the Sun within a large network of stations in various places, in order to eliminate meteorological constraints. The aim is to produce observations 365 days per year, and many times per day. This goal can be achieved with the help of tens of amateurs using a standard spectrograph (SOLEX) and the associated data processing software (INTI) written by Valérie Desnoux [ref 4], which was adapted to the requirements of the BASS2000 solar database for the dissemination of observations. This new collaboration between amateurs and professional astronomers [ref 5] is coordinated at Meudon by Milan Maksimovic.

## 1 – SOLEX OBSERVATIONS FOR THE BASS2000 DATABASE

The BASS2000 solar database (https://bass2000.obspm.fr) accepts CaII H, CaII K and Hα spectral lines. These lines can be easily observed by the SOLEX spectrograph (Figure 1). However, it must be noticed that SOLEX uses lenses for the collimator and the chamber, so that there is some chromatism, and the focus must be adjusted when changing of line. If you use a refractor, the image focus may also be affected. In summary:

- The image focus of the refractor may change between CaII H/K and Hα
- The spectrum focus of the SOLEX changes between CaII H/K and Hα

As a consequence, we recommend to observe either CaII H & K, or Hα, but not both, in order to avoid the refocus of the image on the slit and the refocus of the spectrum. We recommend also Hα in priority (Figures 2 and 3). The examples of Hα monochromatic images of Figure 4 show that SOLEX delivers professional quality data.

The best focal length of the refractor lies in the range 350-480 mm because full disk images are expected by BASS2000. Images should not be cut. The slit of the spectrograph is 4.5 mm long; this corresponds to the maximum possible focal length of 480 mm. If your telescope has a longer focal length, a focal reducer can be incorporated. For instance, the Meudon SOLEX (Figure 5) uses a 625 mm focal length refractor; this is too long, so that the SOLEX is mounted on a 0.67 x optical magnifier which provides the equivalent length of 418 mm.

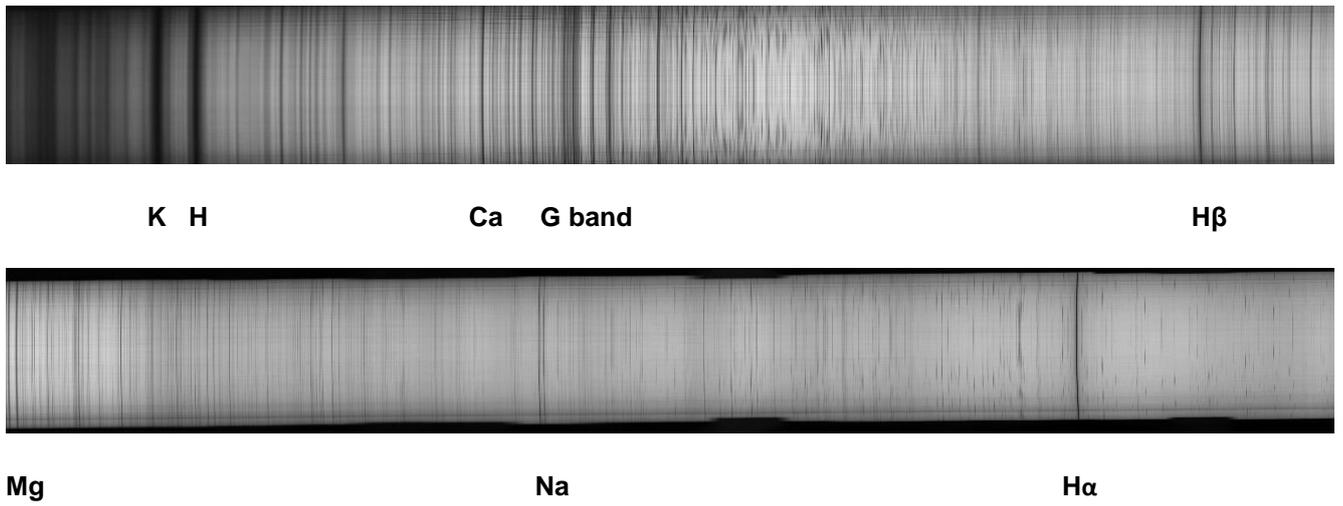

***Figure 1**: SOLEX full spectrum from the violet (CaII H 3968 Å and K 3934 Å lines) to the red (Hα 6563 Å).*

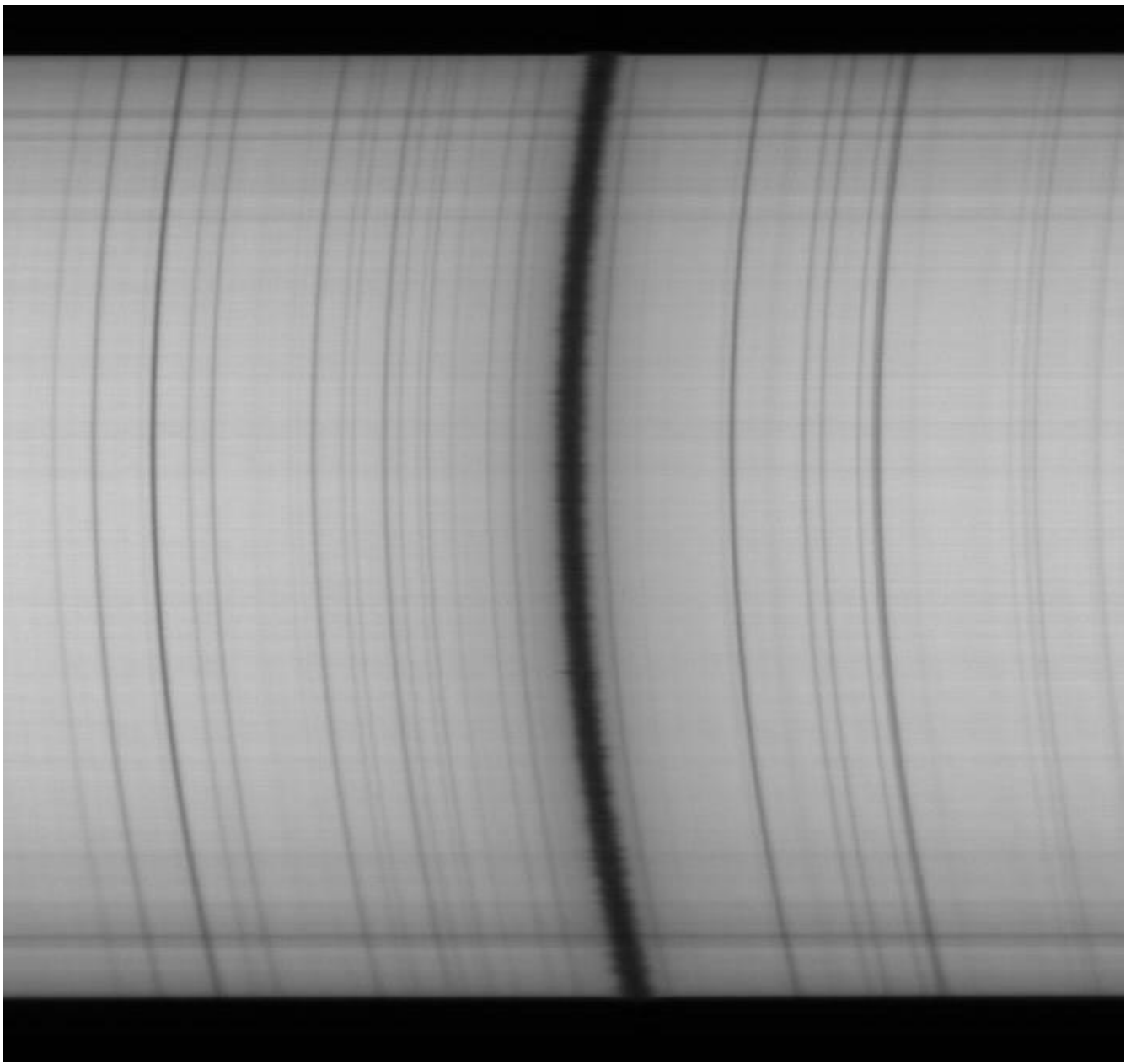

***Figure 2**: The Hα line 6563 Å observed by the SOLEX experiment (the Doppler effect is well visible). Spectral lines are curved and corrected by the INTI software.*

The usual and low cost detector for SOLEX is the SONY IMX178 CMOS sensor (3000 x 2000) with 2.4 µ pixels. This sensor is of course not mandatory, but its characteristics in terms of speed and spectral sampling are well optimized. Bigger pixels are quite possible (maximum 4.8 µ, otherwise the spectral sampling will suffer). The binning 2 x 2 of the IMX178 sensor is often sufficient, it is equivalent to a 4.8 µ pixel camera.

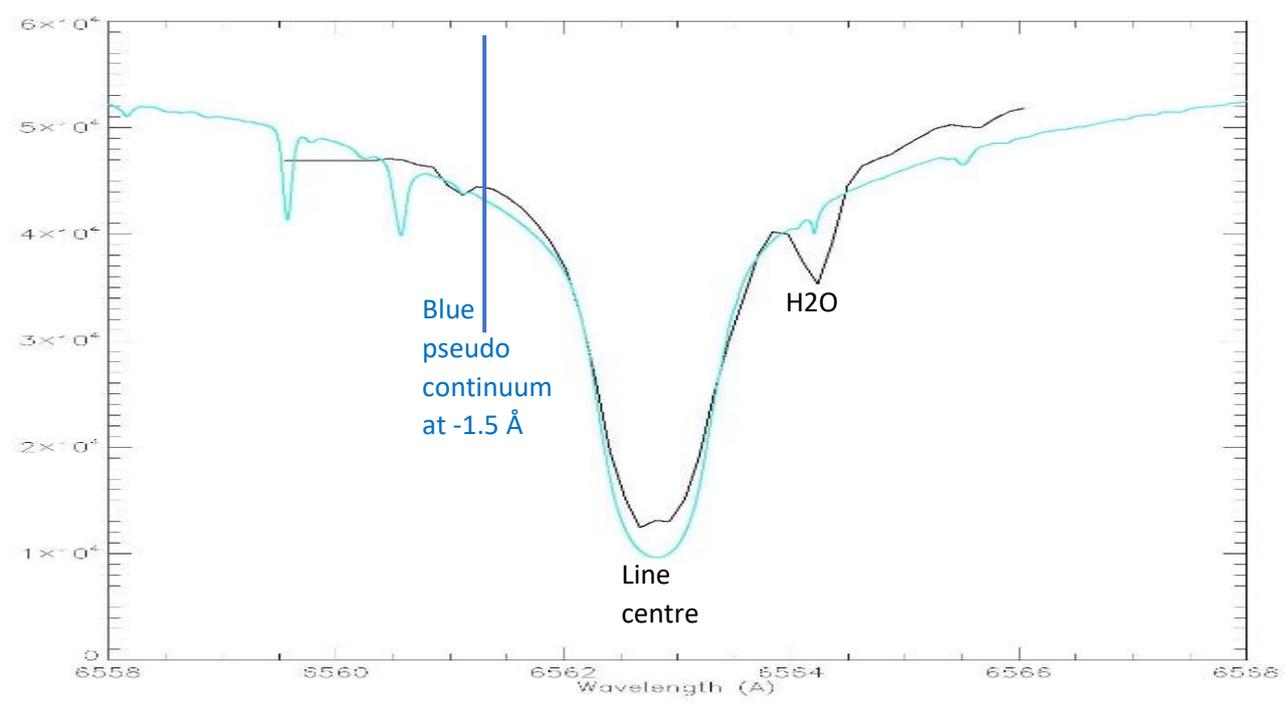

*Figure 3*: *The Hα line at disk centre. Black line: SOLEX observations. Blue line: atlas profile (Delbouille et al, 1973). There is a parasitic terrestrial line (H2O) in the red wing.*

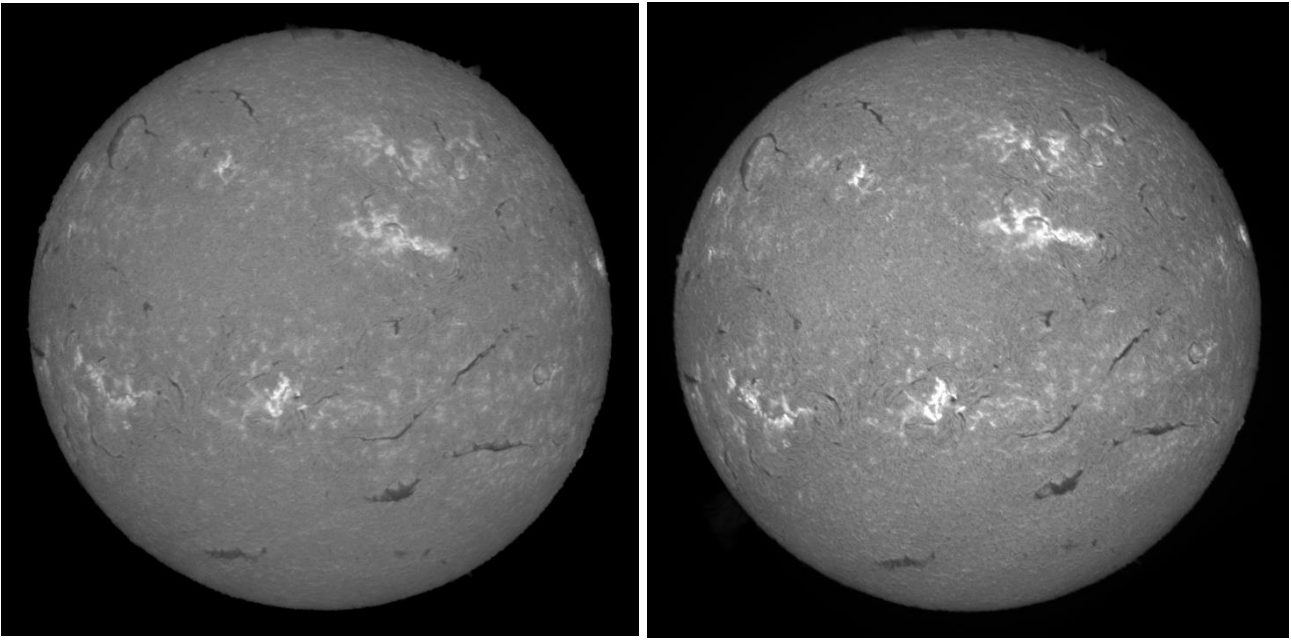

*Figure 4*: *Spectroheliograms in the Hα line 6563 Å observed by the SOLEX experiment (left) and by the Meudon spectroheliograph (right). Example of 9 March 2023.*

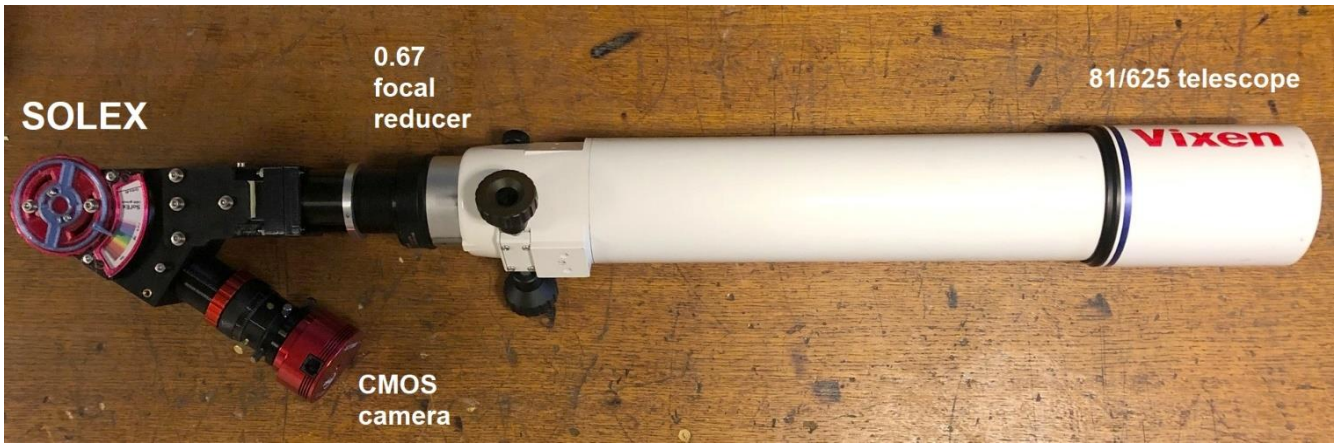

*Figure 5*: The Meudon SOLEX: it uses a 625 mm focal length refractor; the diameter of the solar image exceeds the 4.5 mm slit of the SOLEX and must be reduced. Therefore, we use a 0.67 x focal length magnifier to get an equivalent focal length of 420 mm.

## 2 - FILE TRANSFER TO MEUDON

Observations must be processed by the INTI software **[ref 4]** which guarantees the full compatibility **[ref 6]** with the BASS2000 database. The files must be transferred to the service ftp.obspm.fr using the anonymous FTP protocol in binary mode. FITS files produced by INTI (16 bits) are mandatory; we recommend also to transfer the JPEG files (8 bits) for quick look. Your data will not appear immediately in BASS2000; their compatibility will be checked by a procedure before installation in the database and there is a delay. In case of compatibility problem, you will receive an Email. Your images must be oriented and rotated (via INTI) in order to have the solar North up, the South down, the solar East left and the West right. Images also require to be well centred.

- **Command line transfer (binary mode, MS/DOS, Linux):**
  ftp ftp.obspm.fr
  username: anonymous
  password: your Email address
  cd incoming/cospam
  binary
  put filename

- **With Filezilla (recommended method for personal computers):**
  (software downloadable at https://filezilla-project.org/)
  FTP configuration below:

  | Protocole : | FTP — protocole de transfert de fichiers |
  | Hôte : | ftp.obspm.fr | Port : |
  | Chiffrement : | Connexion FTP simple (non sécurisé) ⚠ |
  | Type d'authentification : | Anonyme |

  Drag and drop INTI files from your directory to the directory incoming/cospam

  Site distant : /incoming/cospam
  - incoming
    - cospam ←
    - ? ec
    - ? HFW
    - ? mrabenanahary

  Nom de fichier
  ..
  _13_05_33Z_SOLEX_APO60FD6_ASI178_EQ6_Ha_20230515_130533.fits
  _13_05_33Z_SOLEX_APO60FD6_ASI178_EQ6_Ha_20230515_130533.jpg
  _13_05_33Z_SOLEX_APO60FD6_ASI178_EQ6_Ha_20230515_130533_protus.jpg

  ← cospam directory with INTI files

# 3 - WAVELENGTHS AND IMAGE NAMES

Image names and FITS keywords (including wavelengths) are those provided by INTI **[ref 4]** and should never be modified; otherwise, the files could be refused by the database in case of compatibility failure. FITS Filenames are terminated by a string containing the line identification, the date and the time (YYYYMMDD = date; HHMMSS = time, the star below represents the beginning of the filename containing the SOLEX identification):

H$\alpha$ line centre (chromosphere)

*_ Ha_YYYYMMDD_HHMMSS.fits

WAVELNTH=         6562.762

H$\alpha$ blue continuum (line centre - 1.5 Å, photosphere)

*_ Ha2cb_YYYYMMDD_HHMMSS.fits

WAVELNTH=         6561.232

CaII K line centre (K3, chromosphere)

*_ CaK_YYYYMMDD_HHMMSS.fits

WAVELNTH=         3933.663

CaII K1v blue wing (line centre -1.5 Å, photosphere)

*_ CaK1v_YYYYMMDD_HHMMSS.fits

WAVELNTH=         3932.163

CaII H line centre (H3, chromosphere)

*_ CaH_YYYYMMDD_HHMMSS.fits

WAVELNTH=         3968.468

CaII H1v blue wing (line centre -1.5 Å, photosphere)

*_ CaH1v_YYYYMMDD_HHMMSS.fits

WAVELNTH=         3966.968

# 4 - IMAGES IN THE CONTINUUM AND IN THE LINE CENTRE

Line centres of CaII H, CaII K and H$\alpha$ are formed in the chromosphere and reveal active regions, dark filaments, prominences at the limb, and bright plages. On the contrary, the wings of the lines are formed below in the photosphere and show sunspots and faculae.

INTI **[ref 6]** is able to compute images in the line centre, in the line wings and in the continuum, when the wavelength shift, with respect to the line centre, is specified in units of pixels.

We recommend for the continuum a shift of 1.5 Å towards the blue (-1.5 Å). If your camera has 2.4 µ pixels, you can use the Figure 6 to get the shift in units of pixels. The figure does not provide the sign: for a shift towards the blue wing (shorter wavelength), it is negative; for a shift towards the red wing (longer wavelength), it is positive. For example, -1.5 Å corresponds to -24 pixels for H$\alpha$. If your camera has 4.8 µ pixels, or 2.4 µ pixels with binning 2, you can use directly the Figure 7 to get the shift in pixels; for example, -1.5 Å corresponds to -12 pixels for H$\alpha$.

If you own a camera with a pixel size of p micrometres, you can still use the Figure 6. Just multiply the result by the factor (2.4/p). For example, for a shift of -1.5 Å from H$\alpha$ centre, Figure 6 indicates -24 pixels. If your camera has a pixel size of 3.6 µ, the shift for your camera will be -24 x (2.4/3.6) = -16 pixels.

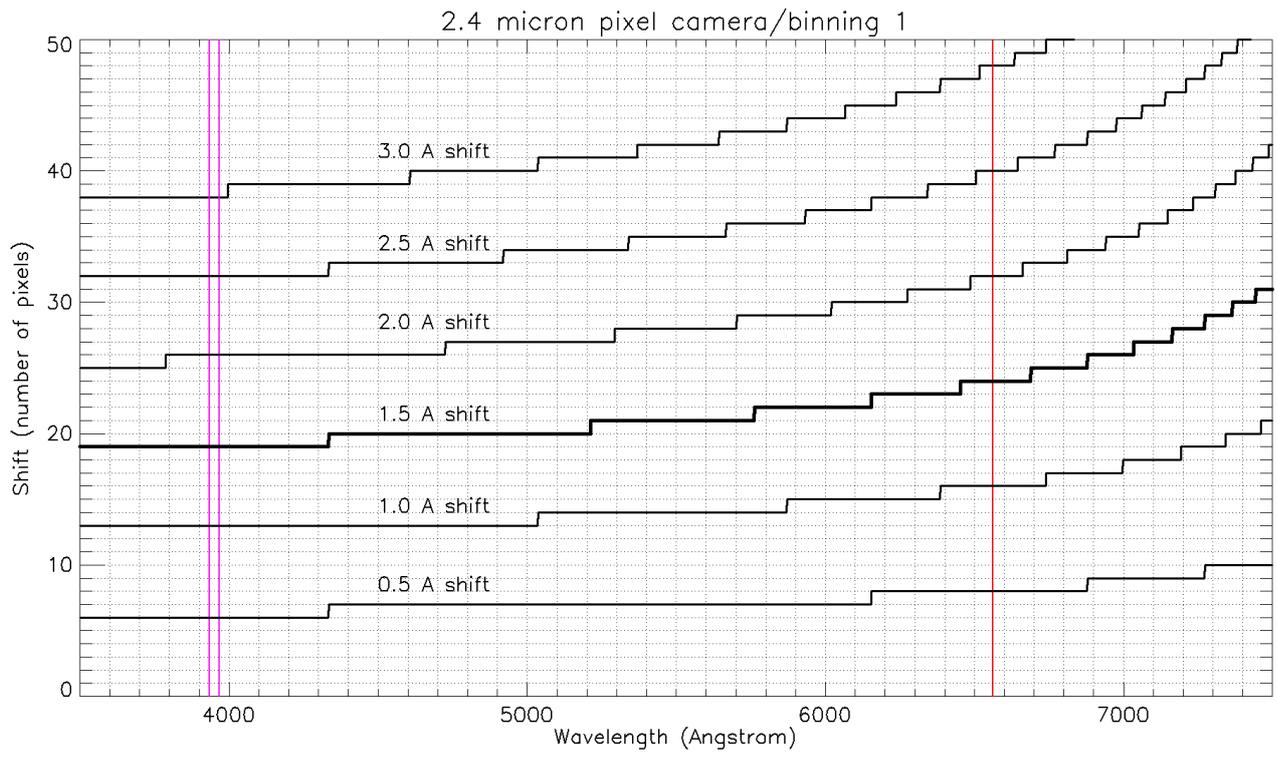

*Figure 6*: *The shift in pixels for a given shift in Angström, as a function of wavelength, for a camera with 2.4 µ pixels. Example: for a shift of -1.5 Å from the Hα line centre (red bar), the shift in pixels is -24; for a shift of -1.5 Å from the CaII H or K line centre (violet bar), the shift in pixels is -19.*

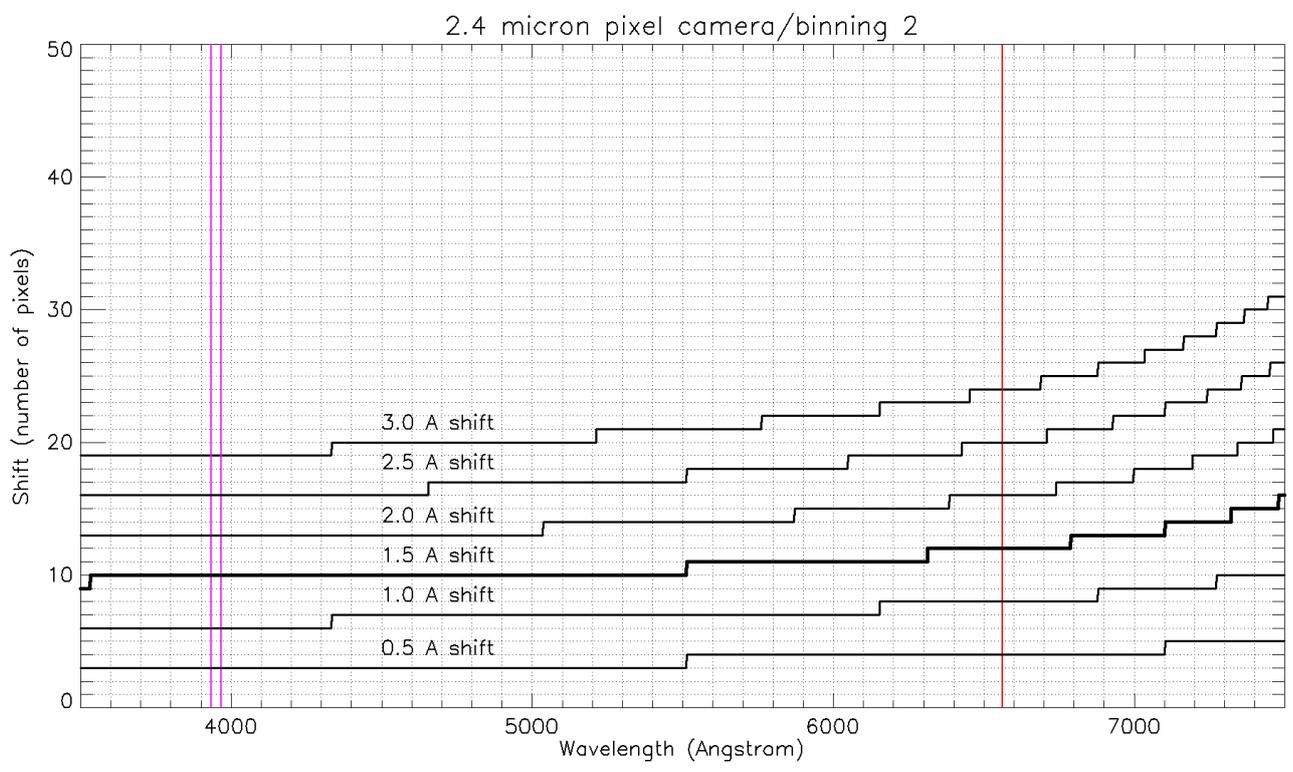

*Figure 7*: *The shift in pixels for a given shift in Angström, as a function of wavelength, for a camera with 4.8 µ pixels (or 2.4 µ pixels with binning 2). Example: for a shift of -1.5 Å from the Hα line centre (red bar), the shift in pixels is -12; for a shift of -1.5 Å from the CaII H or K line centre (violet bar), the shift in pixels is -10.*

# 5 - TECHNICAL INFORMATION: SOLEX CAPABILITIES

The SOLEX is detailed by Buil's website **[ref 1]**. We just provide below some more information. SOLEX can observe many lines of the visible spectrum, because the grating can rotate. Figure 8 presents the theoretical dispersion of the SOLEX with the 2400 grooves/mm grating at interference order 1: it is not constant in wavelength, but increases slowly from 0.03 to 0.04 mm/Å from the violet to the red part of the spectrum. It is not possible to observe lines in the infrared part of the spectrum above 7500 Å. Figure 8 shows also the radius of curvature of spectral lines; it depends on the wavelength, red lines are two times more curved than violet lines. The INTI software **[ref 4]** corrects the curvature before producing monochromatic images.

The spectral theoretical resolution (i.e. the size of smallest details along the line profiles) is displayed in Figure 9 for SOLEX at order 1. It decreases from 0.36 Å for CaII H/K to 0.16 Å for Hα. There is also an order 2 in the ultraviolet part of the spectrum, probably not observable. On this figure, we have reported for comparison the capabilities of Meudon spectroheliograph. As the grating is fixed in a permanent position (no possible rotation), only a few lines are observable. In particular CaII H/K form in interference order 5, while Hα falls in order 3. Meudon spectroheliograph provides optical resolutions of 0.15 Å and 0.25 Å, respectively for CaII H/K and Hα lines. This is quite comparable to SOLEX.

The best camera sampling, according to the Shannon theorem, requires two pixels for the spectral resolution. For example, in Hα, the SOLEX dispersion is 0.038 mm/Å, so that the resolution of 0.16 Å corresponds to 6 μ in the spectrum. One sees that with a pixel smaller than 3 μ, the sampling is perfect.

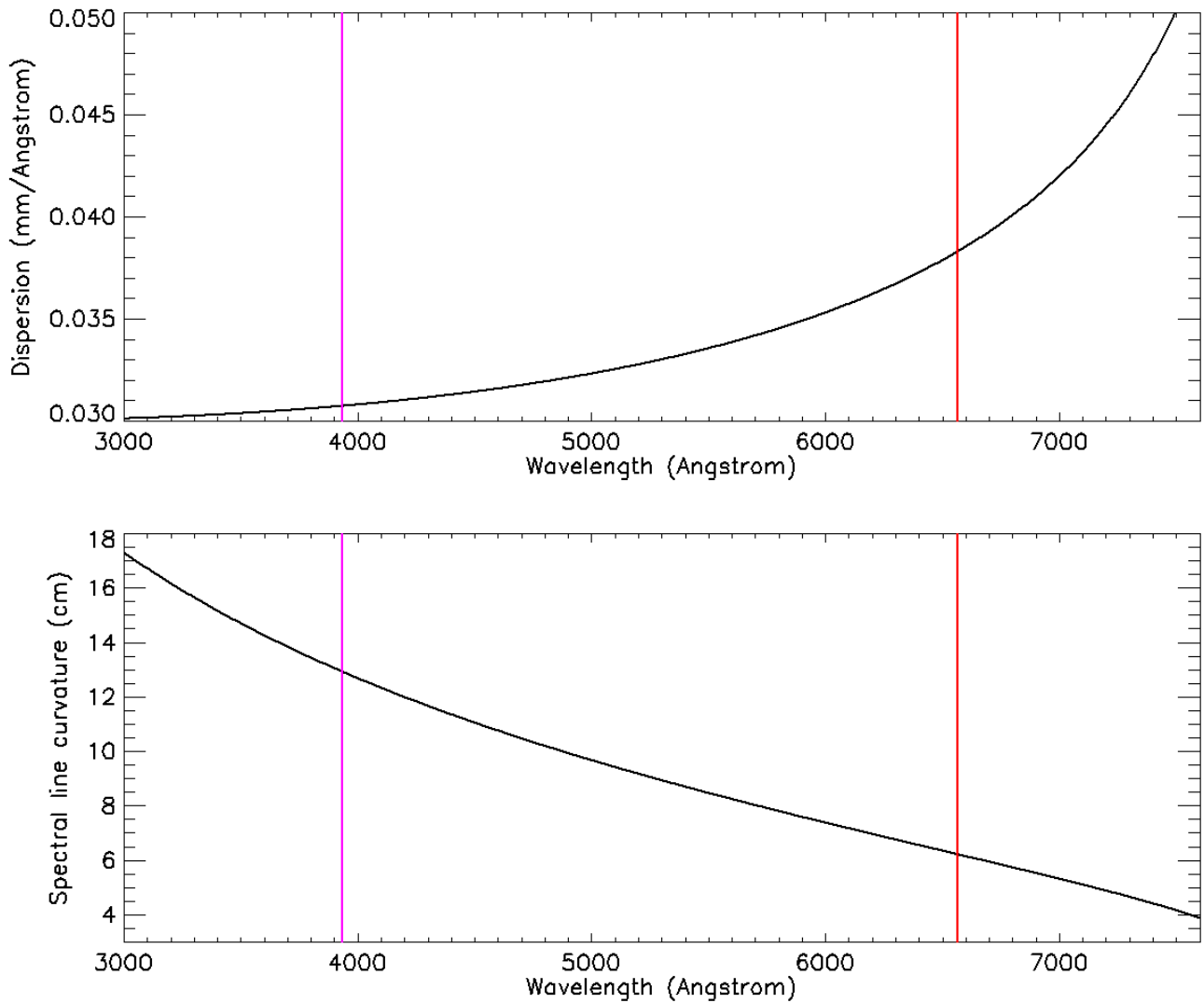

**Figure 8**: Top: The dispersion of SOLEX (in mm/Å) as a function of wavelength. The violet and red bars indicate the position of the CaII H/K and Hα lines. Bottom: the radius of curvature (in cm) of spectral lines as a function of wavelength.

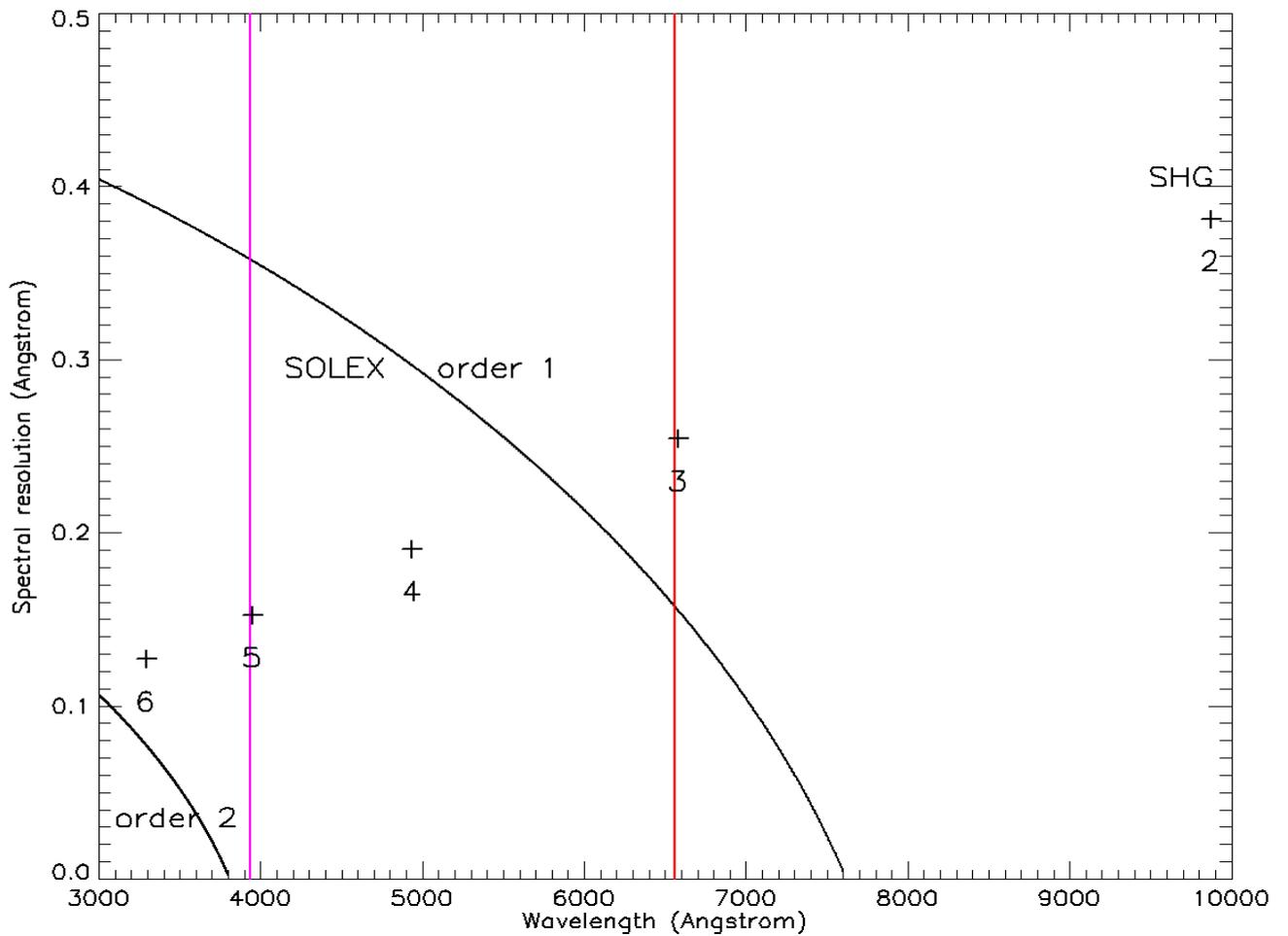

*Figure 9*: Theoretical spectral resolution of SOLEX (in Å) as a function of wavelength. The violet and red bars indicate the CaII H/K and Hα lines. For comparison, the numbers provide the optical resolution of Meudon spectroheliograph in orders 2, 3, 4, 5, 6.

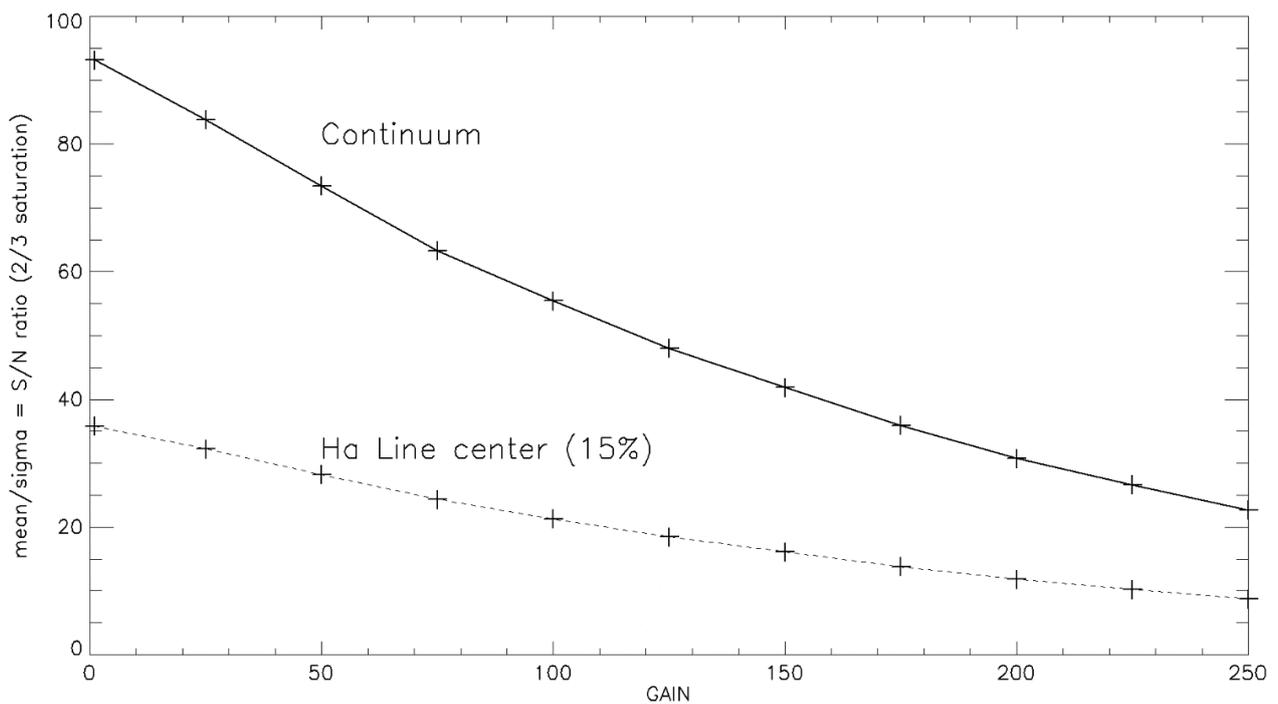

*Figure 10*: The expected Signal to Noise (S/N) ratio for the ZWO IMX178 camera as a function of the gain.

If your camera has pixels of size p, the value of the spectral pixel can be calculated easily using the dispersion d (in mm/Å) of Figure 8. The spectral pixel is (p/d) in Å, where p is expressed in mm. For example, the figure gives d = 0.038 mm/Å for Hα. With p = 2.4 µ = 0.0024 mm pixels, we get for the spectral pixel (0.0024/0.038) = 0.063 Å for Hα line profiles.

The spatial pixel along the slit depends on the equivalent focal length f of the refractor and the magnification γ of the SOLEX (γ is the ratio of the focal length of the chamber, 125 mm, and the collimator, 80 mm, hence γ = 125/80 = 1.56). The diameter D of the Sun at the SOLEX focus is D = 0.0093 (γ f) in mm (with f expressed in mm). For example, f = 420 mm gives D = 6.1 mm for the solar diameter. If p is the pixel size in mm, the spatial pixel in arcsec is simply 1920 (p/D). With p = 2.4 µ = 0.0024 mm and D = 6.1 mm, we get 0.75 arcsec for the spatial pixel. This is much smaller than the slit width of the SOLEX (a few arcsec).

The photon signal to noise ratio (S/N) is an important question. It depends on the full well capacity of each camera pixel measured in electrons. For example, for S = 10000 electrons, N = √S = 100, so that S/N = √S cannot exceed 100. This is typically the maximum S/N ratio for the continuum of lines with the IMX178 sensor. Now, the line cores are much darker, 15% of the continuum for Hα, which means that the number of photo electrons will be 1500 at maximum in the Hα core. Hence, the best photon S/N ratio in Hα with this detector lies around √1500 ≈ 40. The situation deteriorates when the camera gain increases (Figure 10), because S/N decreases with increasing gain. S/N will not exceed 30 for a gain of 50, so that we do not recommend the use of gain values above 50 (at least for ZWO cameras).

## 6 – CONCLUSION: HOW TO CONTRIBUTE

If you want to contribute and provide observations to the BASS2000 database, please fill the form in annex and mail it to Florence Cornu. Then download the INTI software **[ref 4]** and the INTI manual for BASS2000 compatibility **[ref 6]** to process your data. For technical questions concerning your contribution to the database, please contact: Florence.Cornu@obspm.fr

For scientific questions, please contact: Jean-Marie.Malherbe@obspm.fr or Milan.Maksimocic@obspm.fr

By downloading your images to the database, you remain the full owner of your production, but you agree with their freely use by the international community, under the licence CC-BY-NC-SA (Creative Commons, attribution, no commercial use, share alike).

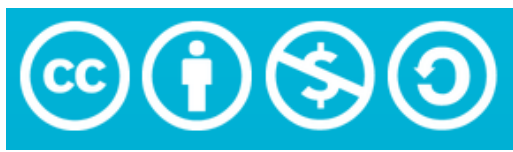 licence CC-BY-NC-SA applying to all images of BASS2000

## 7 - REFERENCES

1. 130 years of spectroheliograms at Paris-Meudon observatories (1892-2022), Malherbe, J.M., 2023, https://doi.org/10.48550/arXiv.2301.11105

2. Optical characteristics and capabilities of the successive versions of Meudon spectroheliograph (1908-2023), Malherbe, J.M., 2023, https://doi.org/10.48550/arXiv.2303.10952

3. The SOLEX instrument, by Christian Buil, http://www.astrosurf.com/solex/sol-ex-presentation-en.html

4. The SOLEX data processing, by Valérie Desnoux, http://valerie.desnoux.free.fr/inti/index.html

5. SOLEX et l'imagerie monochromatique solaire, Buil, C., Malherbe, J.M., Maksimovic, M., 2023, hal-04097987, https://cnrs.hal.science/hal-04097987

6. INTI operator manual for BASS2000 compatibility, Desnoux, V., 2023, http://valerie.desnoux.free.fr/inti/INTI_ProAm_EN.pdf

# ANNEX

| FORM: SOLEX CONTRIBUTOR TO BASS2000 **Please email to : Florence.Cornu@obspm.fr** **with copy to : Milan.Maksimovic@obspm.fr** | | | Internal Ref. | |
|---|---|---|---|---|

| **Observer** | | | | |
|---|---|---|---|---|
| | First name | | | |
| | Name | | | |
| | Email address | | | |
| | Pseudo * | | | |

* your pseudo may appear in BASS2000 with your images. Please choose a pseudo which does not allow to recognize your name; it can be followed by the initials of your country – Please indicate your pseudo in the FITS keyword « OBSERVER » in INTI software

| | | | | |
|---|---|---|---|---|
| I belong to an astronomy association | | ☐ Yes | ☐ No | |
| Please indicate the name of the association | | | | |

**Usual observing site**

| | Country: | | | |
|---|---|---|---|---|
| State, region or district… | Longitude | Elevation | Latitude | |
| | | | | |

| Do you use your instrument in other sites ? | ☐ Never | ☐ Sometimes |
|---|---|---|

**Observed spectral line(s)**

| | | | |
|---|---|---|---|
| Hα line centre (chromosphere, filaments, plages, prominences) | Ha | λ= 6562.762 | ☐ |
| Hα blue continuum (Hα centre – 1.5 Å, photosphere, sunspots) | Ha2cb | λ= 6561.232 | ☐ |
| CaII K line centre (K3) (chromosphere, filaments, plages, prominences) | CaK | λ= 3933.663 | ☐ |
| CaII K blue wing (K1v) (K3 -1.5 Å, photosphere, sunspots, faculae) | CaK1v | λ= 3932.163 | ☐ |
| CaII H line centre (H3) (chromosphere, filaments, plages, prominences) | CaH | λ= 3968.468 | ☐ |
| CaII H blue wing (H1v) (H3 -1.5 Å, photosphere, sunspots, faculae) | CaH1v | λ= 3966.968 | ☐ |
| Other line (please precise) … | | | |

✋ *Important*: If you switch between Hα and CaII H/K, two focus have to be checked (the telescoper image upon the slit and the SOLEX spectrum). For that reason, we do not recommend switching too frequently between Hα and CaII H/K.

**Instrument**

| | | | | |
|---|---|---|---|---|
| Mount type | ☐ Azimuthal | ☐ Equatorial | | |
| Mount model | | | | |
| Telescope/refractor model | | | | |
| Optical formula (diameter / equivalent focal length) | Ø  mm  f  mm | With focal reducer ? | ☐ Yes ☐ No | |
| Camera model | | | | |
| Instrument | ☐ Personal | ☐ Association | | |

| **Processing software** | ☐ INTI | ☐ other : | ? |
|---|---|---|---|
| **Observing frequency** | ☐ Sometimes | ☐ Frequent | ☐ Daily |

**Remarks**